%% file: main.tex
\documentclass[a4paper,11pt]{article}
\usepackage{jcappub} 
\usepackage{lineno}
\usepackage{graphicx} 
\usepackage{physics} 
\usepackage{mathtools} 
\usepackage{hyperref} 
\usepackage[table,xcdraw]{xcolor} 
\usepackage{cleveref}
\usepackage{todonotes}
\usepackage{siunitx}
\usepackage[nice]{nicefrac}
\usepackage{makecell}
\usepackage{subcaption}
\linenumbers

\arxivnumber{} 
\title{Search for quantum decoherence in neutrino oscillations with six detection units of KM3NeT/ORCA\\\line(1,0){420}}

\input{Authors}

\newpage
\abstract{Neutrinos described as an open quantum system may interact with the environment which introduces stochastic perturbations to their quantum phase. This mechanism leads to a loss of coherence along the propagation of the neutrino $-$ a phenomenon commonly referred to as \textit{decoherence} $-$ and ultimately, to a modification of the oscillation probabilities.
Fluctuations in space-time, as envisaged by various theories of quantum gravity, are a potential candidate for a decoherence-inducing environment. Consequently, the search for decoherence provides a rare opportunity to investigate quantum gravitational effects which are usually beyond the reach of current experiments. 
In this work, quantum decoherence effects are searched for in neutrino data collected by the KM3NeT/ORCA detector from January 2020 to November 2021. The analysis focuses on atmospheric neutrinos within the energy range of a few GeV to $100\,\mathrm{GeV}$.
Adopting the open quantum system framework, decoherence is described in a phenomenological manner with the strength of the effect given by the parameters $\Gamma_{21}$ and $\Gamma_{31}$. Following previous studies, a dependence of the type $\Gamma_{ij} \propto (E/E_0)^n$ on the neutrino energy is assumed and the cases $n = -2,-1$ are explored. 
No significant deviation with respect to the standard oscillation hypothesis is observed. Therefore, $90\,\%$ CL upper limits are estimated as $\Gamma_{21} < 4.6\cdot 10^{-21}\,$GeV and $\Gamma_{31} < 8.4\cdot 10^{-21}\,$GeV for $n = -2$ and $\Gamma_{21} < 1.9\cdot 10^{-22}\,$GeV and $\Gamma_{31} < 2.7\cdot 10^{-22}\,$GeV for $n = -1$, respectively.
}

\begin{document}
\nolinenumbers
\maketitle
\flushbottom

\input{Introduction}
\input{Theory}
\input{Detector}
\input{Analysis}
\input{Results}

\input{Conclusion}


\newpage
\acknowledgments

The authors acknowledge the financial support of the funding agencies:
Funds for Scientific Research (FRS-FNRS), Francqui foundation, BAEF foundation.
Czech Science Foundation (GAČR 24-12702S);
Agence Nationale de la Recherche (contract ANR-15-CE31-0020), Centre National de la Recherche Scientifique (CNRS), Commission Europ\'eenne (FEDER fund and Marie Curie Program), LabEx UnivEarthS (ANR-10-LABX-0023 and ANR-18-IDEX-0001), Paris \^Ile-de-France Region, Normandy Region (Alpha, Blue-waves and Neptune), France;
Shota Rustaveli National Science Foundation of Georgia (SRNSFG, FR-22-13708), Georgia;
This work is part of the MuSES project which has received funding from the European Research Council (ERC) under the European Union’s Horizon 2020 Research and Innovation Programme (grant agreement No 101142396).
The General Secretariat of Research and Innovation (GSRI), Greece;
Istituto Nazionale di Fisica Nucleare (INFN) and Ministero dell’Universit{\`a} e della Ricerca (MUR), through PRIN 2022 program (Grant PANTHEON 2022E2J4RK, Next Generation EU) and PON R\&I program (Avviso n. 424 del 28 febbraio 2018, Progetto PACK-PIR01 00021), Italy; IDMAR project Po-Fesr Sicilian Region az. 1.5.1; A. De Benedittis, W. Idrissi Ibnsalih, M. Bendahman, A. Nayerhoda, G. Papalashvili, I. C. Rea, A. Simonelli have been supported by the Italian Ministero dell'Universit{\`a} e della Ricerca (MUR), Progetto CIR01 00021 (Avviso n. 2595 del 24 dicembre 2019); KM3NeT4RR MUR Project National Recovery and Resilience Plan (NRRP), Mission 4 Component 2 Investment 3.1, Funded by the European Union – NextGenerationEU,CUP I57G21000040001, Concession Decree MUR No. n. Prot. 123 del 21/06/2022;
Ministry of Higher Education, Scientific Research and Innovation, Morocco, and the Arab Fund for Economic and Social Development, Kuwait;
Nederlandse organisatie voor Wetenschappelijk Onderzoek (NWO), the Netherlands;
Ministry of Research, Innovation and Digitalisation, Romania;
Slovak Research and Development Agency under Contract No. APVV-22-0413; Ministry of Education, Research, Development and Youth of the Slovak Republic;
MCIN for PID2021-124591NB-C41, -C42, -C43 and PDC2023-145913-I00 funded by MCIN/AEI/10.13039/501100011033 and by “ERDF A way of making Europe”, for ASFAE/2022/014 and ASFAE/2022 /023 with funding from the EU NextGenerationEU (PRTR-C17.I01) and Generalitat Valenciana, for Grant AST22\_6.2 with funding from Consejer\'{\i}a de Universidad, Investigaci\'on e Innovaci\'on and Gobierno de Espa\~na and European Union - NextGenerationEU, for CSIC-INFRA23013 and for CNS2023-144099, Generalitat Valenciana for CIDEGENT/2018/034, /2019/043, /2020/049, /2021/23, for CIDEIG/2023/20, for CIPROM/2023/51 and for GRISOLIAP/2021/192 and EU for MSC/101025085, Spain;
Khalifa University internal grants (ESIG-2023-008 and RIG-2023-070), United Arab Emirates;
The European Union's Horizon 2020 Research and Innovation Programme (ChETEC-INFRA - Project no. 101008324).



\bibliographystyle{JHEP}
\bibliography{biblio.bib}

\end{document}

%% file: Authors.tex
\author[a]{S.~Aiello,}
\author[b,bb]{A.~Albert,}
\author[c]{A.\,R.~Alhebsi,}
\author[d]{M.~Alshamsi,}
\author[e]{S. Alves Garre,}
\author[g,f]{A. Ambrosone,}
\author[h]{F.~Ameli,}
\author[i]{M.~Andre,}
\author[j]{L.~Aphecetche,}
\author[k]{M. Ardid,}
% ...
\affiliation[a]{INFN, Sezione di Catania, (INFN-CT) Via Santa Sofia 64, Catania, 95123 Italy}
\affiliation[b]{Universit{\'e}~de~Strasbourg,~CNRS,~IPHC~UMR~7178,~F-67000~Strasbourg,~France}
\affiliation[c]{Khalifa University, Department of Physics, PO Box 127788, Abu Dhabi, 0000 United Arab Emirates}
\affiliation[d]{Aix~Marseille~Univ,~CNRS/IN2P3,~CPPM,~Marseille,~France}
\affiliation[e]{IFIC - Instituto de F{\'\i}sica Corpuscular (CSIC - Universitat de Val{\`e}ncia), c/Catedr{\'a}tico Jos{\'e} Beltr{\'a}n, 2, 46980 Paterna, Valencia, Spain}
\affiliation[f]{INFN, Sezione di Napoli, Complesso Universitario di Monte S. Angelo, Via Cintia ed. G, Napoli, 80126 Italy}
\affiliation[g]{Universit{\`a} di Napoli ``Federico II'', Dip. Scienze Fisiche ``E. Pancini'', Complesso Universitario di Monte S. Angelo, Via Cintia ed. G, Napoli, 80126 Italy}
\affiliation[h]{INFN, Sezione di Roma, Piazzale Aldo Moro 2, Roma, 00185 Italy}
\affiliation[i]{Universitat Polit{\`e}cnica de Catalunya, Laboratori d'Aplicacions Bioac{\'u}stiques, Centre Tecnol{\`o}gic de Vilanova i la Geltr{\'u}, Avda. Rambla Exposici{\'o}, s/n, Vilanova i la Geltr{\'u}, 08800 Spain}
\affiliation[j]{Subatech, IMT Atlantique, IN2P3-CNRS, Nantes Universit{\'e}, 4 rue Alfred Kastler - La Chantrerie, Nantes, BP 20722 44307 France}
\affiliation[k]{Universitat Polit{\`e}cnica de Val{\`e}ncia, Instituto de Investigaci{\'o}n para la Gesti{\'o}n Integrada de las Zonas Costeras, C/ Paranimf, 1, Gandia, 46730 Spain}
% ...
\affiliation[bb]{Universit{\'e} de Haute Alsace, rue des Fr{\`e}res Lumi{\`e}re, 68093 Mulhouse Cedex, France}
% ...

\emailAdd{Nadja.Lessing@ific.uv.es}
\emailAdd{km3net-pc@km3net.de}

%% file: Introduction.tex
\section{Introduction}\label{sec:intro}
The discovery of neutrino oscillations \cite{TakaakiKajita, ArthurBMcDonald} is one of the first steps towards physics beyond the Standard Model in revealing that neutrinos do have mass, which is not foreseen from the theory. The three-flavour oscillation model has since become well established and the oscillation parameters have been measured by various experiments \cite{NuFit}.
Nevertheless, there are unresolved questions remaining in the neutrino sector concerning for example the Dirac or Majorana nature of neutrinos, the possibility of CP violation, and the determination of the neutrino mass ordering. 
Furthermore, ongoing investigations seek for phenomena that are beyond the predictions of the Standard Model and could modify the oscillation probabilities. These phenomena include neutrino decay, non-standard interactions, and neutrino decoherence.

Neutrino oscillations occur when flavour eigenstates are a coherent superposition of mass eigenstates \cite{Akhmedov_2009}, where each mass eigenstate evolves at a different frequency. As a result, a neutrino initially produced with a specific flavour can be detected with a different flavour after traveling macroscopic distances. The coherency of the neutrino wave is also maintained when neutrinos travel through dense matter. Incoherent inelastic scattering is proportional to $G_F^2$ (where $G_F$ is the Fermi constant) and can safely be neglected. Coherent forward scattering consistently alters the phase velocity of the neutrinos with a term proportional to $G_F$ and to the density of scattering targets, so the interference effect persists.

However, if a neutrino is treated as an open quantum system that interacts with the environment, it can experience stochastic perturbations on its quantum phase, causing the mass eigenstates to lose their coherence. This phenomenon is referred to as \textit{decoherence}.
Fluctuations in the metric of space-time, as anticipated in quantum gravity models, are often suggested as a potential source of decoherence effects \cite{Stuttard_LightCone_2021, Stuttard_Jensen}. Furthermore, a microscopic model for gravitationally induced decoherence has been introduced and compared to phenomenological approaches in a recent work \cite{Alba_theory}. The search for quantum decoherence thus offers a unique opportunity to explore quantum gravity phenomena which are often unobservable with current experimental capabilities.

The signature of neutrino decoherence is a damping of the oscillation amplitude.
Several studies \cite{de_Holanda_2020:2019tuf, IceCubeDeco2023, LongBaselineGomes2023, Solar_and_KAMLAND_Lisi} additionally consider the so-called \textit{relaxation effects} which act on the non-oscillatory terms as part of the decoherence phenomenology.
However, it is important to note that these two effects are distinct phenomena:
pure decoherence can only be probed in experiments that are sensitive to the oscillation terms in the probabilities, whereas relaxation effects can also be tested in scenarios where oscillations are averaged out \cite{oliveira2016quantum:2014jbp}.


Pure relaxation effects have been strongly constrained using solar neutrino data from KamLAND \cite{de_Holanda_2020:2019tuf}.
Pure decoherence has been studied with data from the reactor experiments KamLAND, RENO and Daya Bay \cite{Balieiro_Gomes_2017, De_Romeri_2023}, the accelerator experiments NOvA, MINOS, T2K and OPERA \cite{De_Romeri_2023, Joao_NoVA, Joao_long_basline, LongBaselineGomes2023}, as well as using atmospheric neutrino data from Super-Kamiokande \cite{SuperK_Lisi},  IceCube \cite{Coloma_2018, IceCubeDeco2023}, and IceCube/DeepCore \cite{Coloma_2018}.
A combined search for decoherence and relaxation effects has been conducted with MINOS, T2K \cite{LongBaselineGomes2023} and IceCube \cite{IceCubeDeco2023}.
Furthermore, the impact of decoherence on the precision measurement of standard oscillation parameters has been investigated for future experiments like DUNE, T2HK \cite{Barenboim:2024wdn}, and ESSnuSB \cite{essnusb2024decoherence}.
Since the various experiments are sensitive to different energy regions and oscillation parameters each of them can contribute to constrain a distinctive area of the phase space. 

In this work, decoherence effects are searched for in the data collected with a partial configuration of the KM3NeT/ORCA detector. Following the approach taken in previous studies, a power-law dependence of decoherence effects on the neutrino energy is assumed. 
In \cref{sec:theory} the decoherence model employed in this study is introduced and the effect of decoherence on the oscillation probabilities is illustrated. An overview of the KM3NeT/ORCA detector is provided in \cref{sec:detector}. Information about the data set, the statistical method used to obtain upper limits on the decoherence parameters, and the treatment of systematical uncertainties can be found in \cref{sec:analysis}. Finally, in \cref{sec:results} likelihood curves as well as confidence level contours for decoherence parameters are presented and compared to those obtained with data from other experiments.

%% file: Theory.tex
\section{Theory of neutrino decoherence}\label{sec:theory}

The term \textit{decoherence} refers to the loss of coherence of the neutrino mass eigenstates due to a coupling of the quantum system to a larger environment. This may lead to non-unitary time evolution of the subsystem and introduce irreversible behaviour \cite{Lindblad}. As commonly done for dissipative systems, the time evolution of the neutrino density matrix $\rho$ is described by the Lindblad equation \cite{Stuttard_Jensen, Oliveira_dissipative, Balieiro_Gomes_2017, DUNE, Lindblad}
\begin{equation}\label{eq:TimeEvolution} 
        \frac{d\rho(t)}{dt} = -i[H,\rho(t)] + \mathcal{D}[\rho(t)]\,,
\end{equation}
where $H$ is the Hamiltonian describing standard neutrino oscillations in matter, and the dissipative term $\mathcal{D}[\rho(t)]$ is defined as
\begin{equation}
         \mathcal{D}[\rho(t)] = \frac{1}{2} \sum_{k=1}^{N^2 - 1}  \left(\left[ V_k, \rho(t) V_k^{\dagger} \right]  + \left[ V_k \rho(t) ,V_k^{\dagger} \right] \right)\,,
 \end{equation}
where $V_k$ are general $N \times N$ complex matrices and $N$ the dimension of the Hilbert space of the subsystem. The first term of \cref{eq:TimeEvolution} corresponds to the standard unitary time evolution whereas the second term encodes decoherence effects. For three neutrino families ($N = 3$) the dissipative term can be expanded in the SU(3) basis \cite{Stuttard_Jensen, IceCubeDeco2023, DUNE}
yielding a general form 
 \begin{equation}
 \mathcal{D}[\rho(t)] = (D_{\mu \nu} \rho^\nu)\lambda^\mu\,,
  \end{equation}
where $\lambda^\mu$ are the Gell-Mann matrices, $\rho^\nu$ are coefficients defined by $\rho = \rho_\nu \lambda^\nu$, and $D$ is a $9 \times 9$ symmetric matrix that parameterises decoherence effects\footnote{In this work $D$ is defined in the effective mass basis. Some difficulties of defining $D$ in vacuum and performing an effective matter parameter substitution with decoherence have been addressed in  \cite{Rotation_Vacuum_Matter_issue}.}. 
\newline
Following previous analyses the following physical conditions are imposed:
\begin{itemize}
    \item \textit{Complete positivity} ensures that probabilities of the whole system remain positive at all times \cite{Benatti_2000}.
    This requirement places restrictions on the elements of $D$, rendering them no longer independent \cite{Benatti_2000, De_Romeri_2023, DUNE}. 
    
    \item \textit{Probability conservation} 
    is satisfied if $D_{\mu 0}$ = $D_{0 \nu}$ = 0 \cite{Balieiro_Gomes_2017, DUNE, Buoninfante_2020}.

    \item \textit{Energy conservation} of the subsystem is required even in the presence of matter \cite{DUNE, Coloma_2018}. 

\end{itemize}
The condition of energy conservation implies that relaxation effects which produce a damping in the non-oscillatory terms of the oscillation probabilities 
are disregarded. This is also justified by the expectation that 
the time scale for relaxation effects is much larger than the time scale for pure decoherence \cite{zurek2003decoherence}. 
Furthermore, modifications of the non-oscillatory terms are strongly constrained from solar neutrinos \cite{de_Holanda_2020:2019tuf} which have a significantly larger baseline than the atmospheric neutrinos detected by KM3NeT\footnote{Effects on the oscillatory terms cannot be bound with solar neutrinos since oscillations are averaged out \cite{oliveira2016quantum:2014jbp, Balieiro_Gomes_2017}.}.
\newline
Taking the above into account, decoherence effects can be parameterised by \cite{De_Romeri_2023, DUNE, Balieiro_Gomes_2017, essnusb2024decoherence}
\begin{equation}
    D = - \mathrm{diag}(0, \Gamma_{21}, \Gamma_{21}, 0, \Gamma_{31}, \Gamma_{31}, \Gamma_{32}, \Gamma_{32}, 0)\,.
\end{equation}
The decoherence parameters $\Gamma_{ij}$ are related by \cite{Oliveira_dissipative, DUNE} 
\begin{equation}
    \Gamma_{21} = 2|\vec{a}_3|^2 \geq 0,
\end{equation}
\begin{equation}
    \Gamma_{31} = \frac{1}{2}|\vec{a}_3 + \vec{a}_8|^2 \geq 0,
\end{equation}
\begin{equation}
    \Gamma_{32} = \frac{1}{2}|\vec{a}_3 - \vec{a}_8|^2 \geq 0,
\end{equation}
where the $\vec{a}_p$ are defined by the expansion of the Lindblad master equation in SU(3) with 
\begin{equation}
    \vec{a}_p = (a_p^{(1)}, ..., a_p^{(8)}) \in {\rm I\!R}^8\,,
\end{equation}
\begin{equation}
    V_k = a_p^{(k)} \lambda^p\,.
\end{equation}
Assuming that the angle between $\vec{a}_3$ and $\vec{a}_8$ is zero (as in \cite{Coloma_2018, DUNE}), the following relation can be derived:

\begin{equation}\label{eq:reducedRelationGammas}
    \Gamma_{32} = \Gamma_{31} + \Gamma_{21} - 2\sqrt{\Gamma_{31}\Gamma_{21}}\,,
\end{equation}
resulting in two independent parameters $\Gamma_{21}$ and $\Gamma_{31}$. 
\newline 

In the effective mass basis with matter effects included, standard neutrino oscillations are described by a diagonal Hamiltonian 
\begin{equation}
    H = \frac{1}{2E}
        \begin{pmatrix}
        0 & 0 & 0\\
        0 & \Delta \Tilde{m}_{21}^2 & 0\\
        0 & 0 & \Delta \Tilde{m}_{31}^2
    \end{pmatrix}\,,
\end{equation}
where $E$ is the neutrino energy and $\Delta \Tilde{m}_{ij}^2$ are the effective squared mass differences.
The dissipative term $\mathcal{D}[\rho(t)]$ can be written in terms of the decoherence parameters as \cite{Stuttard_Jensen, Joao_long_basline}
\begin{equation}
    \mathcal{D}[\rho(t)] = -
    \begin{pmatrix}
        0 & \Gamma_{21}\rho_{12}(0) & \Gamma_{31}\rho_{13}(0)\\
        \Gamma_{21}\rho_{21}(0) & 0 & \Gamma_{32}\rho_{23}(0)\\
        \Gamma_{31}\rho_{31}(0) & \Gamma_{32}\rho_{32}(0) & 0
    \end{pmatrix}\,.
\end{equation}
Consequently, the time evolution of the neutrino density matrix as given by \cref{eq:TimeEvolution} has the solution \cite{Balieiro_Gomes_2017, Joao_long_basline}
\begin{equation}
    \rho(t) = 
    \begin{pmatrix}
        \rho_{11}(0) & \rho_{12}(0)e^{-(\Gamma_{21} + i \Tilde{\Delta}_{21})^*t} & \rho_{13}(0) e^{-(\Gamma_{31} + i \Tilde{\Delta}_{31})^*t} \\
        \rho_{21}(0)e^{-(\Gamma_{21} + i \Tilde{\Delta}_{21})t} & \rho_{22}(0) & \rho_{32}(0) e^{-(\Gamma_{32} + i \Tilde{\Delta}_{32})^*t}\\
       \rho_{31}(0)e^{-(\Gamma_{31} + i \Tilde{\Delta}_{31})t} & \rho_{32}(0) e^{-(\Gamma_{32} + i \Tilde{\Delta}_{32})^*t} & \rho_{33}(0)
    \end{pmatrix}\,,
\end{equation}
where $ \Tilde{\Delta}_{ij} = \frac{\Delta \Tilde{m}_{ij}^2}{2E}$. 
\newline
It becomes apparent that the only difference with respect to standard oscillations is the presence of damping terms $e^{-\Gamma_{ij}t}$. 
The probability of a neutrino flavour change, $\nu_\alpha \rightarrow \nu_\beta$, for the decoherence model considered in this work can be found in several papers and reads \cite{DUNE}
\begin{align}
    P(\nu_\alpha \rightarrow \nu_\beta) &= \nonumber
    \delta_{\alpha \beta} - 2 \sum_{i > j} Re(\Tilde{U}_{\beta i} \Tilde{U}^*_{\alpha i} \Tilde{U}_{\alpha j} \Tilde{U}^*_{\beta j}) \\ &+ \nonumber
    2 \sum_{i > j}  Re(\Tilde{U}_{\beta i} \Tilde{U}^*_{\alpha i} \Tilde{U}_{\alpha j} \Tilde{U}^*_{\beta j}) e^{-\Gamma_{ij}L} \cos{\left(\frac{\Delta \Tilde{m}^2_{ij}}{2E}L\right)} \\ &+
    2 \sum_{i > j}  Im(\Tilde{U}_{\beta i} \Tilde{U}^*_{\alpha i} \Tilde{U}_{\alpha j} \Tilde{U}^*_{\beta j}) e^{-\Gamma_{ij}L} \sin{\left(\frac{\Delta \Tilde{m}^2_{ij}}{2E}L\right)}\,,
\end{align}
where $t \approx L$ (in natural units), $L$ being the distance travelled by the neutrino.
As previously emphasised, only oscillatory terms are affected by the damping whereas non-oscillatory terms remain unchanged with respect to standard oscillations. 


In general, decoherence effects may depend on the neutrino energy $E$.
Following previous studies \cite{IceCubeDeco2023, Coloma_2018, De_Romeri_2023} a power-law dependency is assumed 
\begin{equation}
    \Gamma_{ij}(E) = \Gamma_{ij}(E_0) \left( \frac{E}{E_0}\right)^n\,,
\end{equation}
where $E_0 = \SI{1}{GeV}$ is a reference energy and $\Gamma_{ij}(E_0)$ determines the strength of decoherence effects at $E_0$. The index $n$ is typically set to integer values $n = [-2, -1, 0, 1, 2]$, each motivated by different theoretical considerations. 
The $n = -2$ model can be linked to decoherence resulting from light-cone fluctuations \cite{Stuttard_LightCone_2021} and gravitationally induced decoherence\footnote{An exact agreement of a microscopic model and the phenomenological approach was found for vacuum oscillations and $\Gamma_{ij} \propto (\Delta m_{ij}^2)^2$ \cite{Alba_theory}.} \cite{Alba_theory}. 
The $n = -1$ model is the only case which does not violate Lorentz invariance since the damping term follows the same $L/E$ dependence as the oscillation term.

In this work, limits are derived for  $n = -2, -1$ as the KM3NeT/ORCA detector is especially sensitive to modifications of the oscillation probabilities at low energies. 
The current most stringent bounds on the decoherence model considered here are obtained with data from KamLAND, RENO, and T2K \cite{De_Romeri_2023} where their models C, D and E respect the relation between the decoherence parameters stated in \cref{eq:reducedRelationGammas}. 
For $n = 0$ several works find comparable limits 
\cite{De_Romeri_2023, Coloma_2018, IceCubeDeco2023}
whereas for $n = 1, 2$ the best current limits 
are given by IceCube \cite{IceCubeDeco2023}. 

\begin{figure}[h]
    \centering
    \includegraphics[width=0.88\textwidth]{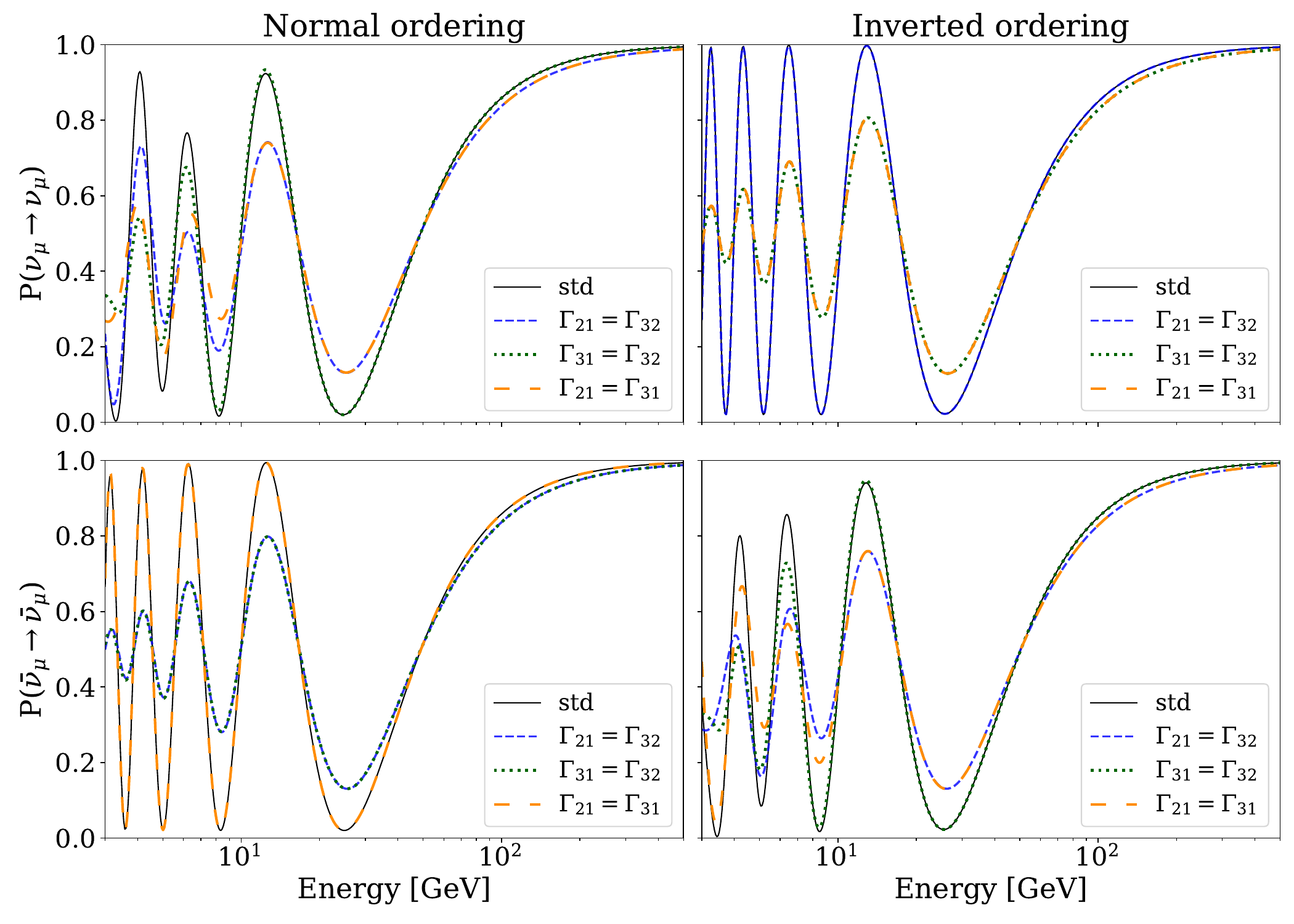}
    \caption{$\nu_\mu$ (upper) and $\bar{\nu}_\mu$ (lower) survival probabilities with normal (left) and inverted (right) ordering for standard oscillations as well as three representative decoherence scenarios, $\Gamma_{21} = \Gamma_{32}$, $\Gamma_{31} = \Gamma_{32}$, and $\Gamma_{21} = \Gamma_{31}$. 
    Decoherence causes a damping of the oscillation amplitude which depends on the mass ordering and whether neutrinos or antineutrinos are considered. The energy dependence shown here corresponds to $n = -1$ and the active decoherence parameters are set to $\Gamma_{ij} = 10^{-22}\,$GeV. }
    \label{fig:1D_probabilities}
\end{figure}

\begin{figure}[h]
    \centering
    \includegraphics[width=0.92\textwidth]{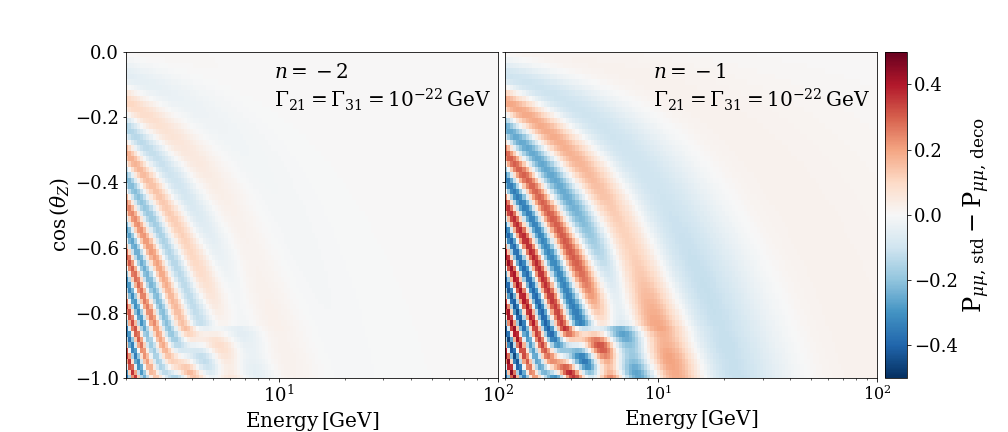}
    \caption{Difference in the $\nu_\mu$ survival probabilities between standard neutrino oscillations, $P_\mathrm{std}$, and assuming decoherence, $P_\mathrm{deco}$, (case $\Gamma_{21} = \Gamma_{31} = 10^{-22}\,\mathrm{GeV}$) as a function of the neutrino energy and zenith angle for $n = -2$ (left) and $n = -1$ (right). The decoherence signal is visible over the whole range of the zenith angle.}
    \label{fig:2D_probabilities}
\end{figure}

Oscillation probabilities are calculated with the open source software OscProb \cite{OscProb}. To account for matter effects the Earth is described by 15 layers of constant matter density based on the Preliminary Reference Earth Model \cite{PREM_model}.
Survival probabilities of up-going $\nu_\mu$ and $\bar{\nu}_\mu$ are shown in \Cref{fig:1D_probabilities} for standard oscillations as well as three representative decoherence scenarios. For the purpose of illustration the values $n = -1$ and $\Gamma_{ij} = 10^{-22}\,$GeV are set for the active decoherence parameters while the remaining decoherence parameter is set to zero in each case. 
This choice of values is in the order of the $90\%\,$CL limit obtained in this work. The damping of the oscillation amplitude in the decoherence scenarios is visible up to $\sim \SI{30}{GeV}$, which is well within the energy range that KM3NeT/ORCA is sensitive to.
Since matter effects are included in the model, the impact of the $\Gamma_{ij}$ parameters on the oscillation probabilities depends on the mass ordering and whether neutrinos or antineutrinos are considered. 
KM3NeT/ORCA detects both neutrinos and antineutrinos, so it is expected to have sensitivity to decoherence for both mass orderings, even if only two of the $\Gamma_{ij}$ are non-zero. In \Cref{fig:2D_probabilities}, the difference between $\nu_\mu$ survival probabilities assuming standard oscillations and assuming decoherence (case $\Gamma_{21} = \Gamma_{31}$) is shown as a function of the neutrino energy and zenith angle. 
The impact of decoherence is visible in the whole range of the zenith angle which means that neutrinos from all directions can contribute to the sensitivity of the detector. 
The standard oscillation parameters are based on the NuFit 5.0 result including Super-Kamiokande data \cite{NuFit}, where the mixing angle has a value of $\theta_{23}=49.2\,^\circ$ for normal ordering.
Since decoherence causes a damping of the oscillation amplitude, its effect can partially be replicated or compensated for by changing the value of the mixing angle. The sensitivity to decoherence is highest for maximal mixing ($\theta_{23} = 45^\circ$) because the difference between the oscillation probability assuming standard oscillations, $P_\mathrm{std}$, and assuming decoherence, $P_\mathrm{deco}$, is largest, making it easier to distinguish between the two.

%% file: Detector.tex
\newpage
\section{The KM3NeT infrastructure}\label{sec:detector}

The KM3NeT research infrastructure comprises two water Cherenkov detectors, referred to as KM3NeT/ARCA (Astroparticle Research with Cosmics in the Abyss) and KM3NeT/ORCA (Oscillation Research with Cosmics in the Abyss), both currently under construction in the Mediterranean Sea \cite{KM3NetLoI}. ARCA is located at a depth of $\SI{3500}{m}$ about $\SI{100}{km}$ offshore from Portopalo di Capo Passero, Sicily, Italy while ORCA is being built about $\SI{40}{km}$ offshore from Toulon, France at a depth of $\SI{2450}{m}$.
The main purpose of ARCA is the search for cosmic neutrinos in the TeV to PeV energy range, whereas ORCA is optimized for 
the measurement of neutrino oscillation properties using atmospheric neutrinos.

Even though the construction is ongoing, KM3NeT has already started taking data using a fraction of the envisaged detector sizes. This is possible due to the modular detector design:
ARCA and ORCA consist of arrays of vertical detection units, each housing 18 digital optical modules \cite{DOM}. These modules are pressure-resistant glass spheres containing 31 photomultiplier tubes (PMTs) each \cite{PMT}, along with their associated readout electronics. The PMTs collect Cherenkov light induced by relativistic charged particles which emerge from neutrino interactions. PMT pulses surpassing a threshold discriminator are digitised and characterised by their starting time and duration. This information is used to reconstruct the initial neutrino direction and energy.


For ORCA the horizontal spacing between detection units ($\sim \SI{20}{m}$) and the vertical distance of digital optical modules ($\sim \SI{9}{m}$) allows for the detection of atmospheric neutrinos with energies as low as a few GeV. This configuration is optimised to enhance the sensitivity to the mass ordering \cite{NMO} and oscillation parameters. Consequently, ORCA can also be used to search for effects which cause a deviation from the oscillation pattern predicted by standard oscillations.


%% file: Analysis.tex
\section{Analysis methods}\label{sec:analysis}
This work uses data collected with a partial configuration of the KM3NeT/ORCA detector comprising six detection units and referred to as ORCA6.
In this section, the event selection and classification as well as the statistical methods of the analysis are explained \cite{ORCA6_oscillations}.

\subsection{Event selection and classification}
The data used in this analysis were taken between January 2020 and November 2021. Only periods characterised by high stability in environmental conditions and data acquisition are utilised. The total livetime is 510 days, corresponding to an exposure of \SI{433}{kton}-years with ORCA6.
Two event topologies are considered in the event reconstruction: \textit{track-like} and \textit{shower-like}. Track-like signatures originate from ${\nu}_\mu$ charged-current (CC) and ${\nu}_\tau$ CC interactions which produce a muon in the final state. Events with an electromagnetic or hadronic shower are induced by ${\nu}_e$ CC and ${\nu}_\tau$ CC interactions, as well as neutral-current (NC) interactions of all neutrino flavours. 
Cuts based on the reconstruction quality reject pure noise events which originate mainly from $^{40}$K decays. The analysis is restricted to up-going events ($\cos(\theta_Z) < 0$) in order to reject atmospheric muons which are a large source of background. A boosted decision tree (BDT) trained on the features of the reconstruction algorithm is employed to discriminate between neutrino-induced signals and poorly reconstructed atmospheric muons. The final sample comprises 5828 observed events with an atmospheric muon contamination below $\SI{2}{\%}$.

A second BDT is used to divide the sample into three distinct classes based on the event topology and reconstruction quality:
i) a high-purity track-like class with negligible atmospheric muon contamination and an estimated $\nu_\mu$ CC purity of $\SI{95}{\%}$, ii)
a low-purity track-like class with $\SI{4}{\%}$ muon contamination and $\SI{90}{\%}$ $\nu_\mu$ CC purity,
and iii) a shower-like class. 
The separation of the track-like events into a high- and a low-purity class improves the sensitivity to standard oscillation parameters due to the isolation of events with a good angular resolution in the high-purity class. 
The track classes contain events with reconstructed energies between $\SI{2}{GeV}$ and $\SI{100}{GeV}$, whereas the shower class ranges from $\SI{2}{GeV}$ to $\SI{1}{TeV}$. 

The data are represented in 2D event histograms of the reconstructed energy, $E_\mathrm{reco}$, and
zenith angle, $\theta_{Z,\mathrm{reco}}$, with a binning scheme that ensures an expectation of at least two events per bin. The detector resolution is described by a response matrix $R(E_\mathrm{true}, \theta_{Z,\mathrm{true}}, E_\mathrm{reco}, \theta_{Z,\mathrm{reco}})$ which is obtained from reconstructed Monte Carlo (MC) events.
The response matrix represents the detection efficiency and reconstruction probability for each bin of true energy, $E_\mathrm{true}$, and zenith angle, $\theta_{Z,\mathrm{true}}$, for each class and flavour. This results in an effective mass of the detector which depends on the interaction type as well as the neutrino flavour and energy. 
More information on the event selection, reconstruction, and classification can be found in \cite{ORCA6_oscillations} which uses the same 
data set as this analysis.

\subsection{Statistical methods}
A likelihood minimisation technique is used to compare 2D reconstructed event histograms to the Monte Carlo expectation, where the values of the oscillation parameters (see  \cref{tab:OscillationParams}) are taken from NuFit 5.0 \cite{NuFit}.
The negative log-likelihood ratio \cite{ORCA6_oscillations}
\small
\begin{align}\label{eq:lnl}
    -2\log{(\mathcal{L}}) = 2
        \sum_{i}
        \,\left[
            (\beta_i N_{i}^\mathrm{mod}
            - N_{i}^\mathrm{dat}) + N_{i}^\mathrm{dat} \log
            \left(
                \frac{N_{i}^\mathrm{dat}}{\beta_i N_{i}^\mathrm{mod}
            }\right)
         \right]
            + \frac{(\beta_i - 1)^2}{\sigma_{\beta i}^2}
            + \sum_k 
        \left(
            \frac{\epsilon_k - \langle\epsilon_k\rangle}{\sigma_{\epsilon k}}
        \right)^2\,,
\end{align}
\normalsize
is minimised over $\theta_{23}$ and $\Delta m_{31}^2$, as well as a set of nuisance parameters $\epsilon_k$ related to model uncertainties. Since ORCA6 is not sensitive to $\theta_{12}$, $\theta_{13}$, $\Delta m^2_{21}$, and $\delta_\mathrm{CP}$ these parameters are fixed in the minimisation.
The first term in \cref{eq:lnl} is derived assuming Poisson distributed reconstructed events with $N_{i}^\mathrm{mod}$ and $N_{i}^\mathrm{dat}$ representing the number of expected events in the model and the number of observed events in data, respectively, per bin of reconstructed energy and zenith angle. The normally distributed $\beta_i$ account for uncertainties due to limited MC statistics following a Barlow and Beeston light method \cite{Barlow:1993dm, ORCA6_oscillations}.
The third term in \cref{eq:lnl} 
constrains model uncertainties by measuring the discrepancy between the observed values $\epsilon_k$ and the expected values $\langle\epsilon_k\rangle$ of the nuisance parameters in units of their standard deviation $\sigma_{\epsilon k}$. Some nuisance parameters  are restricted by a prior while others may vary without constraint around their nominal value (see \cref{tab:priors}). 

\hfill

The model uncertainties can be categorised into three groups:
\begin{enumerate}
    \item Uncertainties in the atmospheric neutrino flux are addressed by nuisance parameters which vary the ratios between muon and electron neutrinos as well as neutrinos and antineutrinos. The proportion of horizontal and up-going neutrinos can be adjusted by varying the shape of the flux in dependence of the zenith angle. The variation of the spectral index produces a tilt in the energy distribution of the flux.

    \item Uncertainties in the neutrino cross section and selection efficiency are taken into account by adjusting the total number of neutrinos as well as the relative number of neutrino events in the shower and high-purity track class. Additionally, scaling factors are applied to the number of NC events and the number of $\nu_\tau$ CC events. A relative normalisation accounts for approximations used in the light simulation of high energy neutrino events ($E_\mathrm{true} > \SI{500}{GeV}$ for CC events, $E_\mathrm{true} > \SI{100}{GeV}$ for NC events). The absolute number of atmospheric muons may vary without constraint.

    \item Uncertainties in the optical properties of water, PMT efficiencies, and the hadronic shower propagation are modelled by an absolute energy scale of the detector which may introduce a shift in the true energy of the response function.
    
\end{enumerate}

\begin{table}[h]
\centering
\begin{tabular}{|c|c|c|c|}
\hline
\textbf{Parameter} & \textbf{Value NO} & \textbf{Value IO} & \textbf{Treatment} \\ \hline
$\Delta m^2_{31}\,[\mathrm{eV}^2]$ & $2.517 \cdot 10^{-3}$  & $-2.424 \cdot 10^{-3}$ & free \\ \hline
$\Delta m^2_{21}\,[\mathrm{eV}^2]$ & $7.42 \cdot 10^{-5}$ & $7.42 \cdot 10^{-5}$ & fixed \\ \hline
$\theta_{12}$\,[$^\circ$] & 33.44 & 33.45 & fixed \\ \hline
$\theta_{13}$\,[$^\circ$] & 8.57 & 8.60 & fixed \\ \hline
$\theta_{23}$\,[$^\circ$] & 49.2 & 49.3 & free \\ \hline
$\delta_\mathrm{CP}$\,[$^\circ$] & 197 & 282 & fixed \\ \hline
\end{tabular}
\caption{Oscillation parameters taken from NuFit 5.0 \cite{NuFit}.}
\label{tab:OscillationParams}
\end{table} 

\begin{table}[h]
\centering
    \begin{tabular}{|c|c|}
    \hline
    \textbf{Parameter} & \textbf{Uncertainty} \\ \hline
    $\nicefrac{(\nu_\mu + \bar{\nu}_\mu)}{(\nu_e + \bar{\nu}_e)}$ 
    & $\pm 2\,\%$ \\ \hline
    $\nicefrac{\nu_e}{\bar{\nu}_e}$ & $\pm 7\,\%$ \\ \hline
    $\nicefrac{\nu_{\mu}}{\bar{\nu}_{\mu}}$ & $\pm 5\,\%$ \\ \hline
    $\nicefrac{\nu_\mathrm{hor}}{\nu_\mathrm{up}}$  & $\pm 2\,\%$ \\ \hline
    Spectral index & $\pm 10\,\%$ \\ \hline
    $f_\mathrm{all}$ & unconstrained \\ \hline
    $f_\mathrm{S}$ & unconstrained \\ \hline
    $f_\mathrm{HPT}$ & unconstrained \\ \hline
    $f_\mathrm{NC}$ & $\pm 20\,\%$ \\ \hline
    $f_{\tau\mathrm{CC}}$ & $\pm 20\,\%$ \\ \hline
    $f_\mathrm{HE}$ & $\pm 50\,\%$ \\ \hline
    $f_\mu$ & unconstrained \\ \hline
    $E_s$ & $\pm 9\,\%$ \\ \hline
    \end{tabular}
    \caption{Nuisance parameters with their uncertainties.}\label{tab:priors}
\end{table}

A more detailed description of the systematic parameters and their modelling can be found in \cite{ORCA6_oscillations}.


%% file: Results.tex
\newpage
\section{Results}\label{sec:results} 

No significant deviation was found with respect to standard oscillations as the best-fit values of $\Gamma_{21}$ and $\Gamma_{31}$ are consistent with zero. 
In order to derive upper limits, confidence intervals are obtained from a scan of the log-likelihood ratio $-2\log{(\mathcal{L}_\mathrm{deco}/\mathcal{L}_\mathrm{bf})} = -2 \Delta \log \mathcal{L}$, where $\mathcal{L}_\mathrm{deco}$ is computed for a set of points in the decoherence phase space and $\mathcal{L}_\mathrm{bf}$ is the likelihood at the global best fit.

\subsection{Likelihood ratio scans}

\begin{figure}[h]
    \centering
    \includegraphics[width=\textwidth]{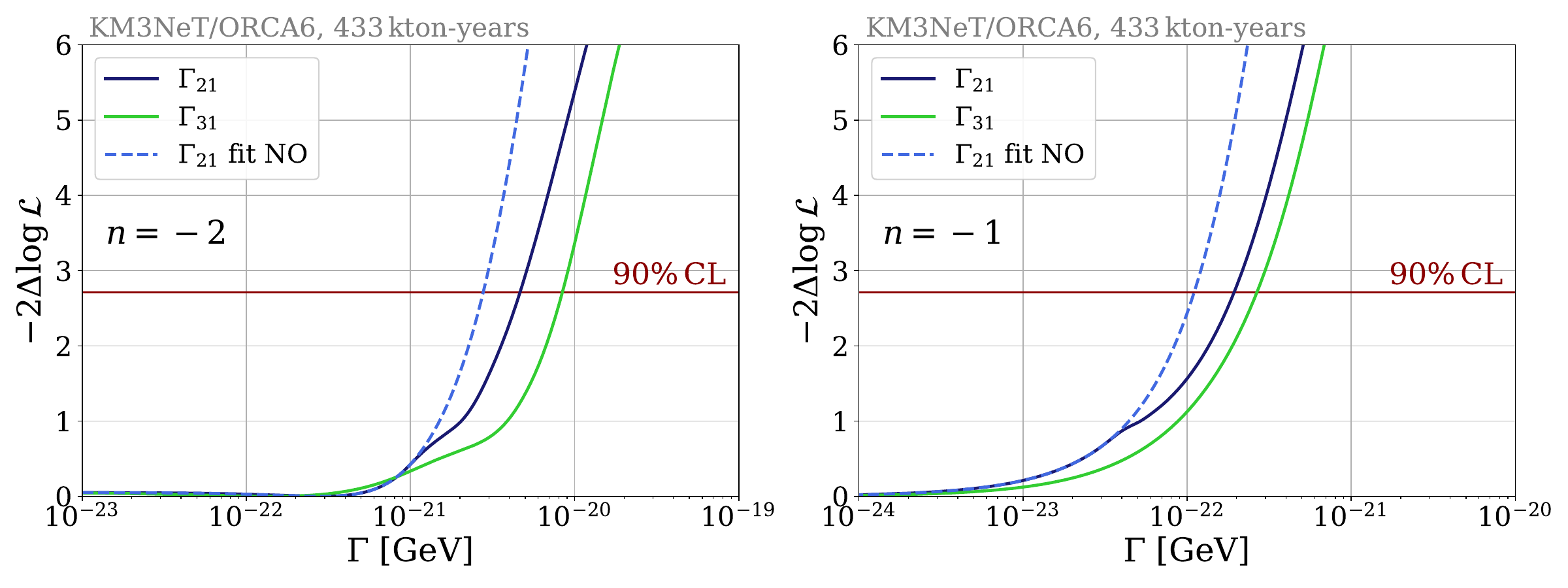}
    \caption{Log-likelihood ratio with respect to the global best fit for $n = -2$ (left) and $n = -1$ (right) as a function of the decoherence parameters $\Gamma_{21}$ (blue) and $\Gamma_{31}$ (green). The solid lines were obtained fitting both, NO and IO and keeping the overall best fit. The dashed line was obtained assuming NO which demonstrates that IO is preferred for large values of $\Gamma_{21}$.}
    \label{fig:chi2Curve}
\end{figure}

The log-likelihood ratio $-2 \Delta \log \mathcal{L}$ as a function of the decoherence parameters can be seen in \Cref{fig:chi2Curve} for both energy dependencies $n = -2$ (left) and $n = -1$ (right). The solid blue curve corresponds to a scan over $\Gamma_{21}$ and the green curve to a scan over $\Gamma_{31}$. In each scan, the remaining decoherence parameter is left free in the fit which allows for non-zero values of all three $\Gamma_{ij}$. 
Each minimisation uses eight starting points, $\theta_{23} = \{40^\circ, 50^\circ\}$, $E_s = \{0.95, 1.05 \}$ and $\Delta m^2_{31} = \{-2.428\cdot 10^{-3}, 2.517\cdot 10^{-3} \}\,\mathrm{eV}^2$, where $\theta_{23}$ is restricted to the lower/upper octant, the energy scale to below/above one, and $\Delta m^2_{31}$ to the corresponding mass ordering.
The overall best fit is represented by solid lines whereas the dashed line additionally shows the result for normal ordering (NO).
For $\Gamma_{31}$ normal ordering always gives the smaller log-likelihood ratio.
Regarding $\Gamma_{21}$, normal ordering is preferred for small values, and inverted ordering (IO) is preferred for large values.

\begin{figure}[h]
\centering
\begin{subfigure}{.49\textwidth}
  \centering
  \includegraphics[width=1\linewidth]{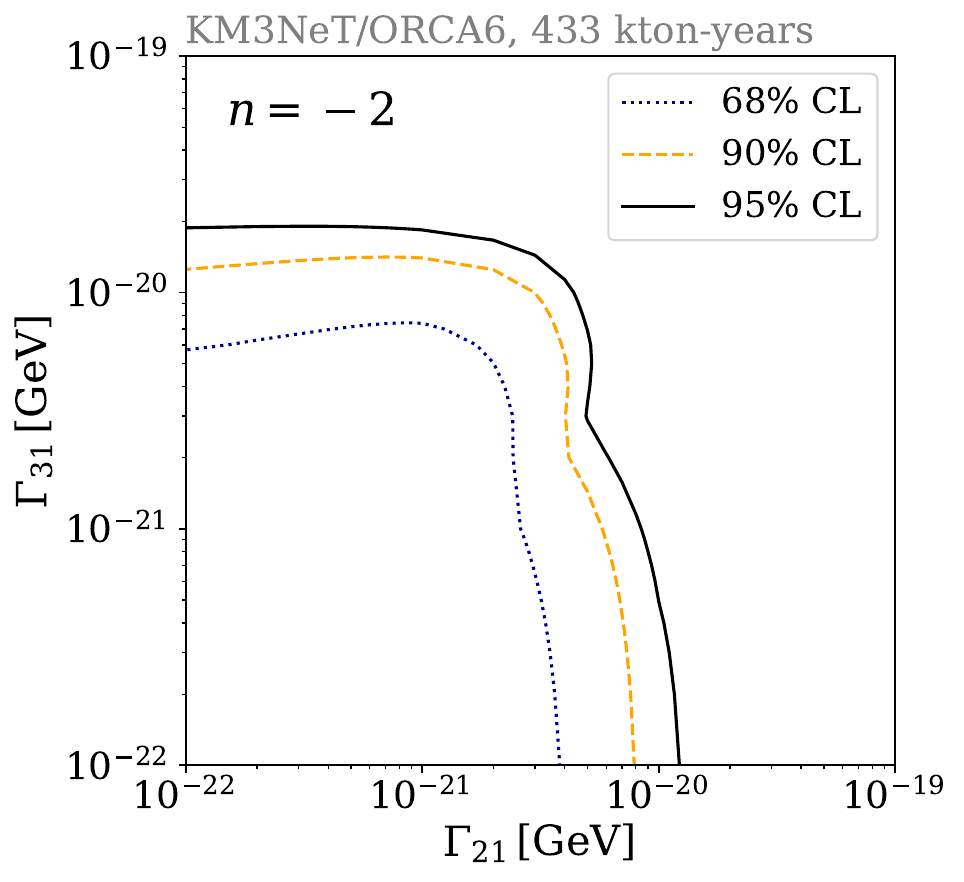}
  \label{fig:sub1}
\end{subfigure}%
\begin{subfigure}{.49\textwidth}
  \centering
  \includegraphics[width=1\linewidth]{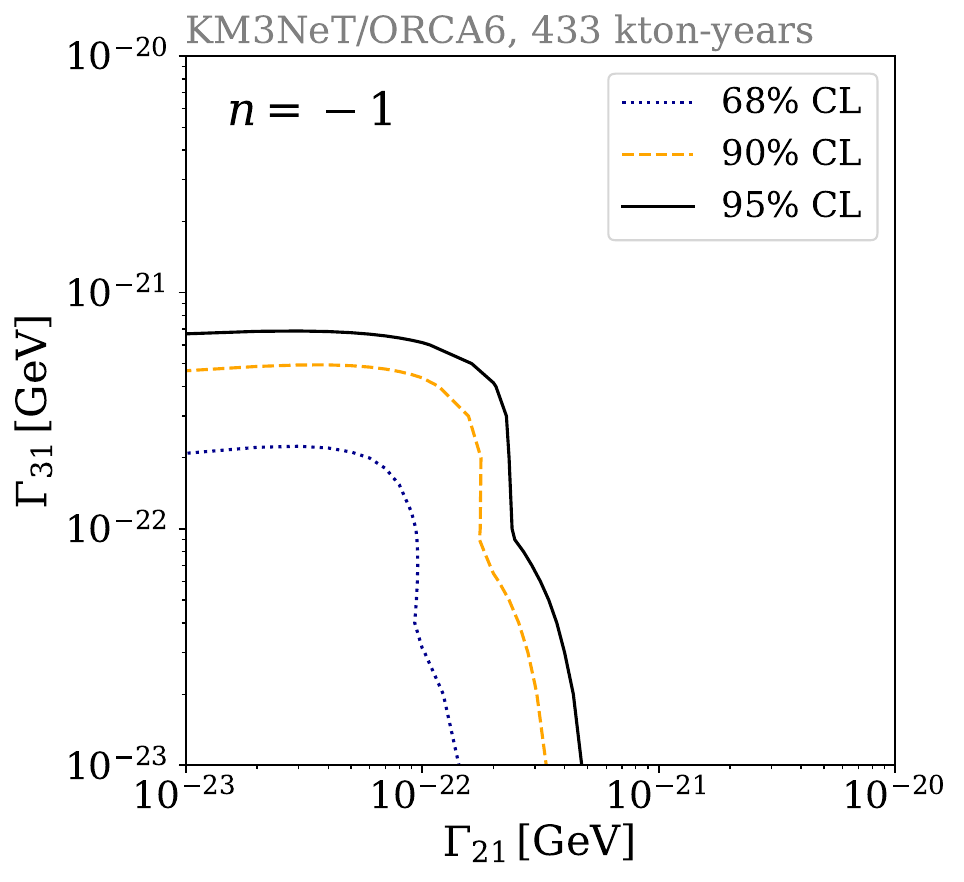}
  \label{fig:sub2}
\end{subfigure}
\caption{Confidence level contours of $\Gamma_{31}$ and $\Gamma_{21}$ for $n = -2$ (left) and $n = -1$ (right). The upper right part of the parameter space is excluded for each model at the corresponding CL. For $\Gamma_{21} = \Gamma_{31}$ the log-likelihood ratio is small as $\Gamma_{32}$ does not contribute to decoherence effects.}
\label{fig:FullContour}
\end{figure}

\begin{figure}[h]
\centering
\begin{subfigure}{.49\textwidth}
  \centering
  \includegraphics[width=1\linewidth]{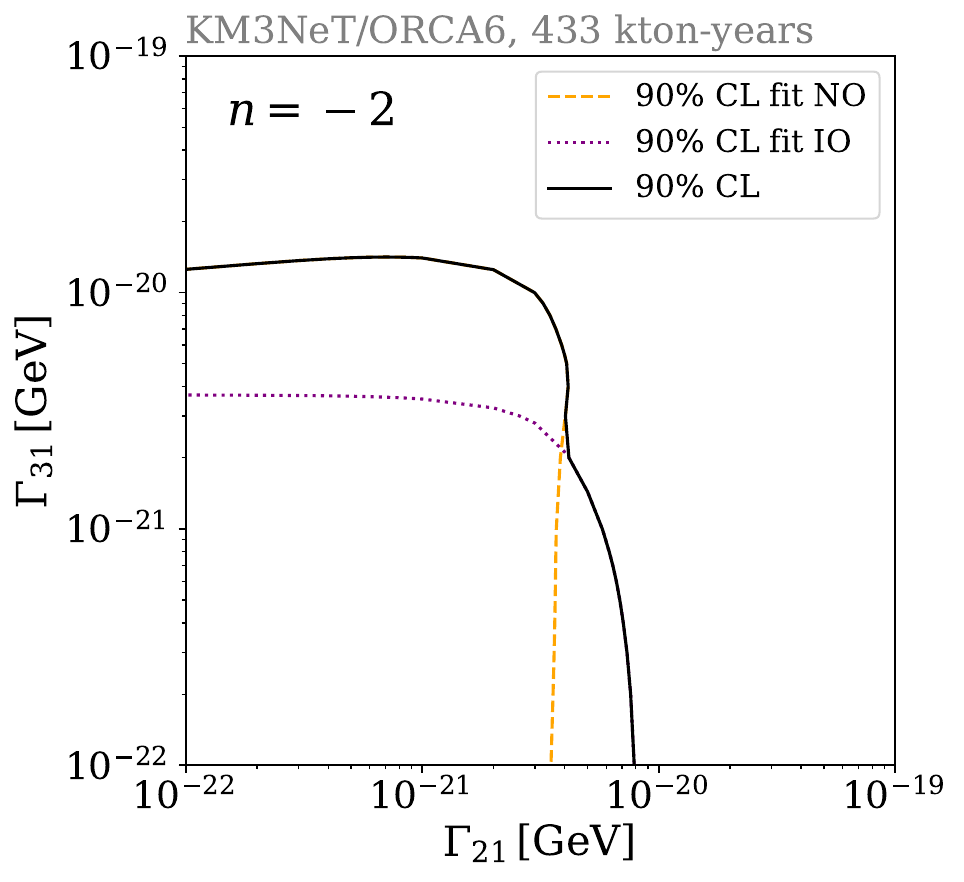}
  \label{fig:sub1}
\end{subfigure}%
\begin{subfigure}{.49\textwidth}
  \centering
  \includegraphics[width=1\linewidth]{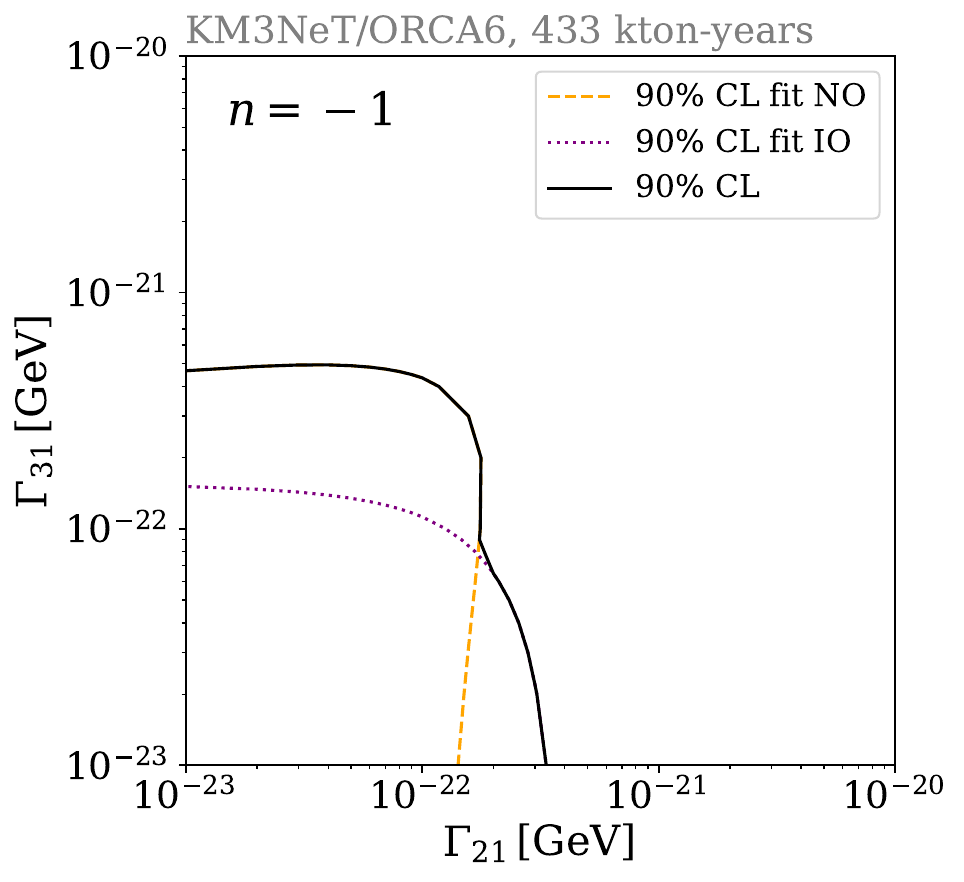}
  \label{fig:sub2}
\end{subfigure}
\caption{$90\%\,$CL contours of $\Gamma_{31}$ and $\Gamma_{21}$ assuming NO (orange dashed), assuming IO (purple dotted), and fitting both mass orderings (black solid) for $n = -2$ (left) and $n = -1$ (right). IO is preferred in the minimisation when decoherence effects are dominated by $\Gamma_{21}$.}
\label{fig:MHContour}
\end{figure}

Confidence level contours which simultaneously constrain $\Gamma_{21}$ and $\Gamma_{31}$ are shown in \Cref{fig:FullContour} for both energy dependencies. The upper right part of the parameter space is excluded at the corresponding CL. The log-likelihood ratio is small for $\Gamma_{21} = \Gamma_{31}$ since from \cref{eq:reducedRelationGammas} $\Gamma_{21} = \Gamma_{31} \Rightarrow \Gamma_{32} = 0$, implying that $\Gamma_{32}$ does not contribute to decoherence effects. 
The shape of the contours can be understood from \Cref{fig:MHContour} which shows the contour at $90\%\,\mathrm{CL}$ assuming NO (orange, dashed) or IO (purple, dotted) respectively, along with the best fit ordering (black, solid). As in the one-dimensional profiles, IO is preferred when decoherence effects are dominated by $\Gamma_{21}$.

\subsection{Upper limits and comparison to previous studies}

Upper limits are reported in \cref{tab:Limits_ORCA6} for both energy dependencies $n = -2,-1$ as well as both mass orderings at the $90\%\,\mathrm{CL}$ and $95\%\,\mathrm{CL}$, assuming that the log-likelihood ratio asymptotically approaches a $\chi^2$-distribution. 
The limits on the scenario $\Gamma_{21} = \Gamma_{31}$ are obtained from the confidence level contours and reported for comparison with previous studies. 

\begin{table}[h]
\centering
\begin{tabular}{|c|cc|cc|}
\hline
  &
  \multicolumn{4}{c|}{Upper limits [GeV]} \\ \hline
  &
  \multicolumn{2}{c|}{$n = -2$} &
  \multicolumn{2}{c|}{$n = -1$} \\ \hline
ORCA6, $90\,\%$CL &
  \multicolumn{1}{c|}{NO} &
  IO &
  \multicolumn{1}{c|}{NO} &
  IO \\ \hline
$\Gamma_{21}$ &
  \multicolumn{1}{c|}{$2.8 \cdot 10^{-21}$} &
  $\mathbf{4.6 \cdot 10^{-21}}$ &
  \multicolumn{1}{c|}{$1.1 \cdot 10^{-22}$} &
  $\mathbf{1.9 \cdot 10^{-22}}$ \\ \hline
$\Gamma_{31}$ &
  \multicolumn{1}{c|}{$\mathbf{8.4 \cdot 10^{-21}}$} &
  $2.2 \cdot 10^{-21}$ &
  \multicolumn{1}{c|}{$\mathbf{2.7 \cdot 10^{-22}}$} &
  $0.8 \cdot 10^{-22}$ \\ \hline
$\Gamma_{21} = \Gamma_{31}$ &
  \multicolumn{1}{c|}{$\mathbf{4.1 \cdot 10^{-21}}$} &
  $2.9 \cdot 10^{-21}$ &
  \multicolumn{1}{c|}{$\mathbf{1.8 \cdot 10^{-22}}$} &
  $1.1\cdot 10^{-22}$ \\ \hline
ORCA6, $95\,\%$CL &
  \multicolumn{1}{c|}{NO} &
  IO &
  \multicolumn{1}{c|}{NO} &
  IO \\ \hline
$\Gamma_{21}$ &
  \multicolumn{1}{c|}{$3.7 \cdot 10^{-21}$} &
  $\mathbf{6.9 \cdot 10^{-21}}$ &
  \multicolumn{1}{c|}{$1.6 \cdot 10^{-22}$} &
  $\mathbf{3.0 \cdot 10^{-22}}$ \\ \hline
$\Gamma_{31}$ &
  \multicolumn{1}{c|}{$\mathbf{11.7 \cdot 10^{-21}}$} &
  $3.2 \cdot 10^{-21}$ &
  \multicolumn{1}{c|}{$\mathbf{4.2 \cdot 10^{-22}}$} &
  $1.3 \cdot 10^{-22}$ \\ \hline
$\Gamma_{21} = \Gamma_{31}$ &
  \multicolumn{1}{c|}{$\mathbf{5.2 \cdot 10^{-21}}$} &
  $3.6 \cdot 10^{-21}$ &
  \multicolumn{1}{c|}{$\mathbf{2.3 \cdot 10^{-22}}$} &
  $1.4 \cdot 10^{-22}$ \\ \hline
\end{tabular}
\caption{Upper limits on the decoherence parameters at $90\,\%$\,CL and $95\,\%$\,CL  for the energy dependencies $n = -2, -1$ for NO and IO. The more conservative limit for both mass orderings is highlighted in bold. 
}
\label{tab:Limits_ORCA6}
\end{table}


\begin{table}[h]
\centering
\begin{tabular}{|c|cc|cc|}
\hline
  &
  \multicolumn{4}{c|}{Upper limits [GeV]} \\ \hline
  &
  \multicolumn{2}{c|}{$n = -2$} &
  \multicolumn{2}{c|}{$n = -1$} \\ \hline
Reported in \cite{De_Romeri_2023}, $90\,\%$\,CL& \multicolumn{2}{c|}{} & \multicolumn{2}{c|}{}  \\ \hline
$\Gamma_{21} = \Gamma_{32}$ & \multicolumn{2}{c|}{$7.9 \cdot 10^{-27}$ (KL)} & \multicolumn{2}{c|}{$1.8 \cdot 10^{-24}$ (KL)}  \\ \hline
$\Gamma_{31} = \Gamma_{32}$ & \multicolumn{2}{c|}{$6.9 \cdot 10^{-25}$ (R)}  & \multicolumn{2}{c|}{$2.1 \cdot 10^{-23}$ (T2K)} \\ \hline
$\Gamma_{21} = \Gamma_{31}$ & \multicolumn{2}{c|}{$7.9 \cdot 10^{-27}$ (KL)} & \multicolumn{2}{c|}{$1.8 \cdot 10^{-24}$ (KL)}  \\ \hline

Reported in \cite{Coloma_2018},  $95\,\%$CL& \multicolumn{1}{c|}{NO} & IO & \multicolumn{1}{c|}{NO} & IO \\ \hline
$\Gamma_{21} = \Gamma_{32}$ & \multicolumn{1}{c|}{$7.5 \cdot 10^{-21}$} & $\mathbf{5.0 \cdot 10^{-20}}$ & \multicolumn{1}{c|}{$3.5 \cdot 10^{-22}$} & $\mathbf{2.3 \cdot 10^{-21}}$ \\ \hline
$\Gamma_{31} = \Gamma_{32}$ & \multicolumn{1}{c|}{$\mathbf{4.3 \cdot 10^{-20}}$} & $1.4 \cdot 10^{-20}$ & \multicolumn{1}{c|}{$\mathbf{2.0 \cdot 10^{-21}}$} & $5.8 \cdot 10^{-22}$ \\ \hline
$\Gamma_{21} = \Gamma_{31}$ & \multicolumn{1}{c|}{$\mathbf{1.2 \cdot 10^{-20}}$} & $8.3 \cdot 10^{-21}$ & \multicolumn{1}{c|}{$\mathbf{5.4 \cdot 10^{-22}}$} & $3.6 \cdot 10^{-22}$ \\ \hline
\end{tabular}
\caption{Upper limits at $90\,\%$\,CL for three representative cases of decoherence  using data from KamLAND (KL), RENO (R), and T2K \cite{De_Romeri_2023} and limits at $95\,\%$\,CL using three years of DeepCore data \cite{Coloma_2018}. The limits reported for DeepCore display the same dependency on the mass ordering as observed for ORCA6. In \cite{De_Romeri_2023} no distinction is made between the orderings.}
\label{tab:Limits_DC_Vale}
\end{table}



A comparison with previous studies is not straightforward due to the diverse approaches that have been employed to set limits on decoherence. Additionally, some previous works \cite{LongBaselineGomes2023, IceCubeDeco2023} neglect the relation between the decoherence parameters outlined in  \cref{eq:reducedRelationGammas}, 
allowing for scenarios such as $(\Gamma_{21} \neq 0, \Gamma_{31} = \Gamma_{32} = 0)$ or $(\Gamma_{21} = \Gamma_{31} = \Gamma_{32} \neq 0)$ which are not possible within the framework considered here. As already mentioned, the upper limits provided in this work are determined for individual decoherence parameters $\Gamma_{ij}$ while profiling over the remaining parameter\footnote{The same approach was taken in a sensitivity study for DUNE \cite{DUNE}.}. However, we found that when computing the log-likelihood scan for $\Gamma_{21}$ ($\Gamma_{31}$) the fitted value of the remaining parameter $\Gamma_{31}$ ($\Gamma_{21}$) tends to be significantly smaller than the parameter of interest. Therefore, limits on $\Gamma_{21}$ ($\Gamma_{31}$) are approximately comparable to limits on $\Gamma_{21} = \Gamma_{32}$ ($\Gamma_{31} = \Gamma_{32}$) reported in \cite{Coloma_2018, De_Romeri_2023} and listed in \cref{tab:Limits_DC_Vale}. Strong bounds were obtained using data from KamLAND, RENO and T2K \cite{De_Romeri_2023}. These experiments can access lower energies than ORCA and are therefore especially sensitive to decoherence for negative values of $n$. The limits  derived using three years of DeepCore data \cite{Coloma_2018} are comparable to those of ORCA6.

\subsection{\boldmath Signature of decoherence at $95\%\,$CL}
\begin{figure}[h] 
\centering
  \includegraphics[width=0.98\linewidth]{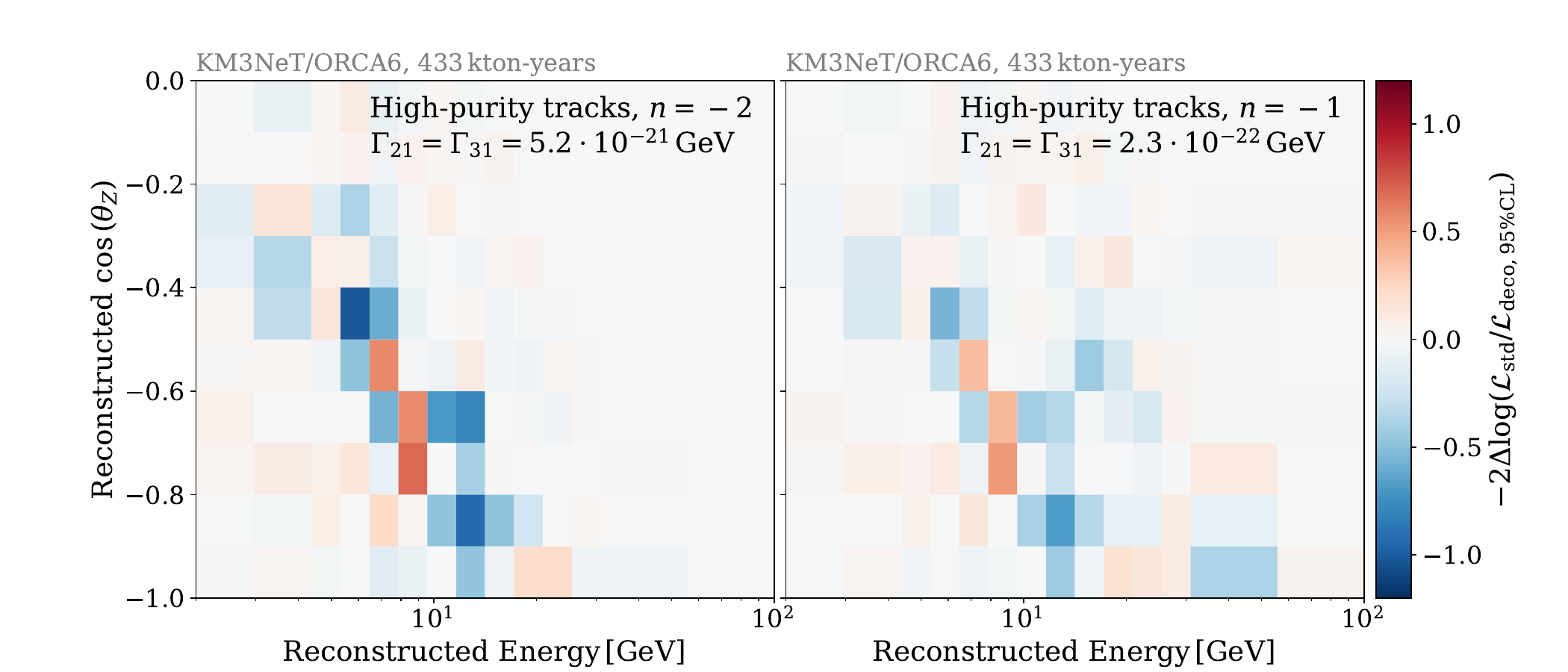}
\caption{Log-likelihood ratio between decoherence and standard oscillations for $n = -2$ (left) and $n = -1$ (right) as a function of the reconstructed energy and zenith angle for the \textit{high-purity track} class. The decoherence parameters are fixed at the respective $95\%\,$CL upper limit with $\Gamma_{21} = \Gamma_{31}$. Negative values (blue) correspond to a better fit of standard oscillations than decoherence.}
\label{fig:Chi2Maps}
\end{figure}

\Cref{fig:Chi2Maps} depicts the log-likelihood ratio between the best fit for standard oscillations and decoherence ($\Gamma_{21} = \Gamma_{31}$) at the corresponding $95\%\,$CL upper limit for $n = -2,-1$ assuming normal ordering. The ratio is presented as a function of the reconstructed energy and zenith angle for the \textit{high-purity track} class, which has the largest contribution to the sensitivity.
Negative values (blue) indicate a preference for standard oscillations. 
The distribution of bins contributing to the sensitivity follows the expected pattern, reflecting the difference in  oscillation probabilities $\mathrm{P}_\mathrm{std} - \mathrm{P}_\mathrm{deco}$ shown in \Cref{fig:2D_probabilities}. As anticipated, up-going neutrinos coming from all directions contribute to the sensitivity. Events with $\cos(\theta_Z) < -0.4$ have a larger contribution since the \textit{high-purity track} class contains more events coming from below the detector than events coming from the horizon ($\cos(\theta_Z) = 0)$. 

\begin{figure}[h]
\centering
  \includegraphics[width=0.65\linewidth]{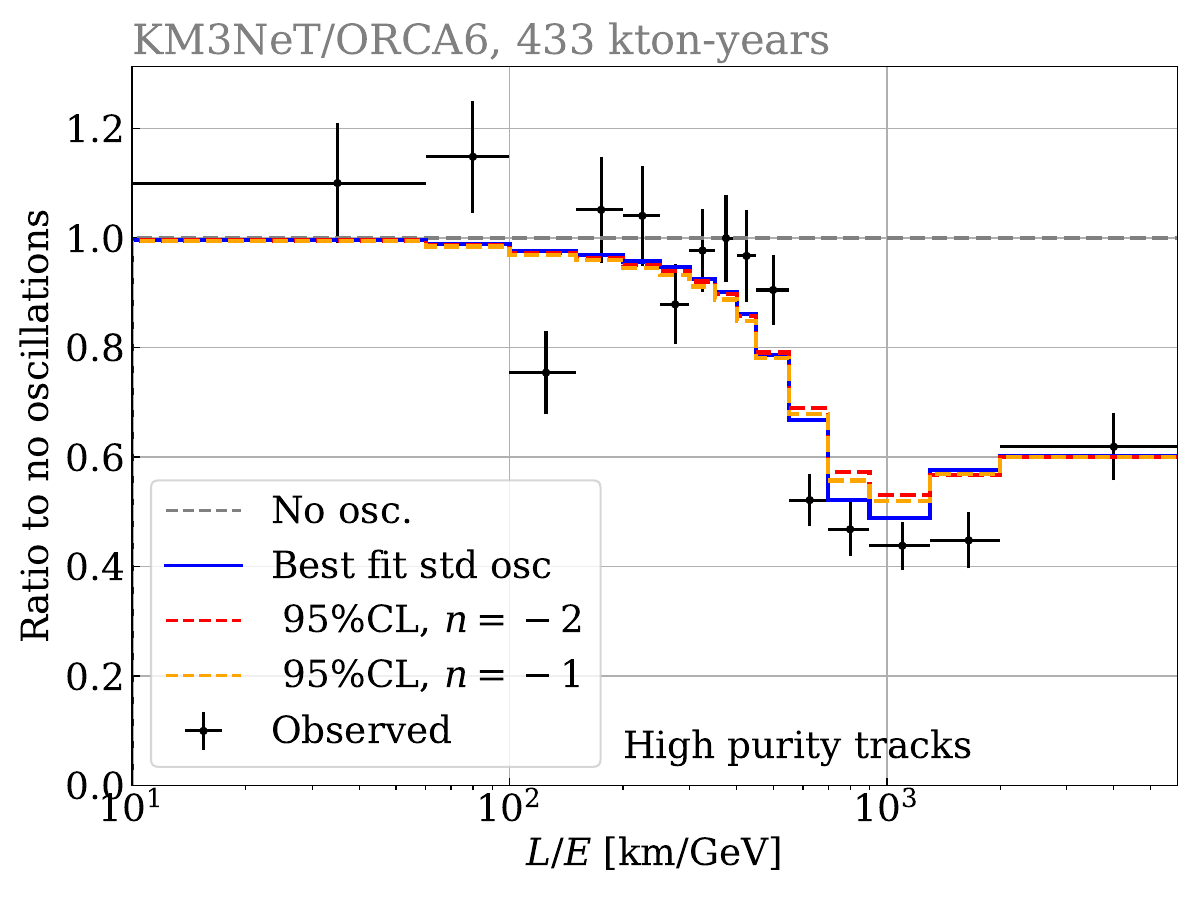}
\caption{Ratio of events with respect to the no oscillations hypothesis as a function of $L/E$ for the standard oscillations best fit and the decoherence models $n = -2,-1$ with $\Gamma_{21} = \Gamma_{31}$ fixed at their respective $95\%\,$CL upper limit. A deviation with respect to standard oscillations can be seen at $L/E \approx 10^3\,\frac{\mathrm{km}}{\mathrm{GeV}}$.
}
\label{fig:LoE}
\end{figure}

To visualise the effect of decoherence, the fit result is transformed into the baseline energy ratio, $L/E$, 
and normalised to the no oscillations hypothesis. This is shown in \Cref{fig:LoE} for standard oscillations as well as the decoherence models $n = -2,-1$ with $\Gamma_{21} = \Gamma_{31}$ fixed at their respective $95\%\,$CL upper limit assuming NO. 
It can be seen that decoherence effects in ORCA6 are mainly visible in two bins at $L/E \approx 10^3\,\mathrm{km}/\mathrm{GeV}$.

%% file: Conclusion.tex
\newpage
\section{Conclusions}
Searches for quantum decoherence effects in neutrino oscillations have been carried out using a $433\,$kton-years data set from the KM3NeT/ORCA detector, in a partial configuration with six detection units.
The strength of decoherence effects was parameterised in terms of the decoherence parameters $\Gamma_{21}$, $\Gamma_{31}$
assuming a power-law dependency on the neutrino energy $\Gamma_{ij} \propto (E/E_0)^n$.
Upper limits on the parameters $\Gamma_{21}$ and $\Gamma_{31}$ have been derived for the cases $n = -2,-1$ showing that the decoherence sensitivity of KM3NeT/ORCA depends on the neutrino mass ordering. Upper limits are therefore reported for both, normal and inverted ordering. The results are comparable to bounds reported for IceCube/DeepCore and display the same dependency on the mass ordering.